# Virtual Immersive Reality for Stated Preference Travel Behaviour Experiments: A Case Study of Autonomous Vehicles on Urban Roads

Bilal Farooq$^{\psi}$

Assistant Professor, Laboratory of Innovations in Transportation (LITrans)
Ryerson University, Toronto, Canada
phone: +1 416 979 5000 x 6456
fax: +1 416 979 5122
bilal.farooq@ryerson.ca

Elisabetta Cherchi

Professor, School of Civil Engineering and Geosciences,
Newcastle University, Newcastle, UK
phone: +44 191 208 3501
elisabetta.cherchi@ncl.ac.uk

Anae Sobhani

Postdoctoral Researcher, Transport and Logistics Group
TU Delft, Delft, the Netherlands
phone: +31 15 27 88384
a.sobhani@tudelft.nl

6153 Words + 4 Figures + 1 Tables = 7403

$^{\psi}$ Corresponding author

**ABSTRACT**

Stated preference experiments have been known to suffer from the lack of realism. This issue is particularly visible when the scenario doesn't have a well understood prior reference e.g. in case of the autonomous vehicles related scenarios. We present Virtual Immersive Reality Environment (VIRE) that is capable of developing highly realistic, immersive, and interactive choice scenario. We demonstrate the use of VIRE in the pedestrian preferences related to autonomous vehicles and associated infrastructure changes on urban streets of Montréal. The results are compared with predominantly used approaches i.e. text-only and visual aid. We show that VIRE results in better understanding of the scenario and consistent results.

**Keywords:** virtual reality, augmented reality, stated preference experiment, behavioural analysis, autonomous vehicle, pedestrian behaviour

## INTRODUCTION

Stated Preference (SP) experiments involve presenting the respondents with carefully created hypothetical scenario in order to efficiently assess their choice preferences *(1)*. They are considered to be a strong tool used in transportation to evaluate behavioural responses–especially in cases of the alternatives that do not exist in the current market. However, SP surveys have been criticized for the lack of realism i.e. not realistically representing the actual choice situation *(2)*. This is particularly true in the case of highly innovative services and technologies, which respondents have never heard of or have very little understanding of.

Connected and Autonomous Vehicles (CAV) are expected to become part of the urban vehicular fleet in coming years. CAV technology has the potential to revolutionize mobility and profoundly impact land use, regional economics, individual behaviour, culture, and society *(3)*. Recently, web-based SP survey techniques have been used to study the questions like demand of AV, willingness-to-pay, and adaption rates *(3, 4)*. However, these studies primarily relied on the respondent's mental image of the CAV and their perception of what CAV may or may not be able to do. The respondents understanding of the CAV technology and capabilities may be limited and heterogeneous. The lack of experience and hence of knowledge of the product may have a major impact on the estimation of individuals' preferences. The degree of realism possible in a text-heavy web-based SP survey is also limited. Furthermore, for more detailed questions, like understanding pedestrian's perception, the lack of actual experience in web-based technologies raises questions about the relevance and quality of their generated datasets.

The use of visualization and simulation in SP experiments has shown to improve realism *(2)*. Patterson et al. concluded that the use of visuals is highly beneficial when the they convey accurate description of the choice scenarios. Due to the recent technological development in the Head Mounted Display (HMD), virtual and augmented reality is becoming more realistic, representative of the reality, and accessible. Various consumer HMD are now available e.g. *Oculus Rift, HTC Vive, Microsoft Hololens*. This creates a great opportunity for the use of virtual and augmented reality to develop realistic and representative SP experiments–especially in the situations where the mental image about a technology or service is very limited and tasks are complicated to be expressed textually.

In this paper, we present the Virtual Immersive Reality Environment (VIRE) platform for developing a range of transportation SP experiments that are highly realistic, immersive, interactive, and have the capability to collect additional information about the motion, orientation, and oral reactions (Ohs and Ahs) of the respondents. VIRE not only addresses the realism issue in SP experiments, but it also gives us the ability to design more detailed experiments that can investigate dimensions like mental processes in choice experiment, without putting any extra burden on the respondent. The information provided to the respondent can also be well-controlled and manipulated. We demonstrate the capabilities of VIRE in a choice situation where the preferences of pedestrians in the context of autonomous vehicles on urban roads are investigated. In this controlled experiment, respondents physically cross a virtual road intersection and interact with the autonomous vehicles. We also compared the VIRE based SP experiment with text-only and standard visualization based SP experiments.

Rest of the paper is organized as follows. Next section presents a brief overview on the use of visualization in SP experiments and the general use of SP experiments in studying the behavioural dimension of the forthcoming technologies. We then present the VIRE platform and its features.

The experiment design is presented that involves pedestrian walking in a virtual intersection with autonomous vehicles. We then present the results and their analysis. The paper is concluded with remarks on VIRE capabilities and possible uses.

**BACKGROUND**

The SP experiments research community proactively explores new tools in order to make the choice tasks and thus responses more realistic. Patterson et al. *(2)* conducted location choice experiments using text and gaming engine based animations. They concluded that the combination of visual and textual information increased the processing of information by the respondents. Cherchi and Hensher *(1)* pointed out the need for visual and engaging tools e.g. eyes tracking, virtual reality, and simulators to add value to behavioural relevance and reduce hypothetical bias. However, due to the amount of work needed to actually develop these tools, very little progress has been made in terms of the main stream usage of more immersive tools in SP experiments.

This fact is evident from the recent literature on the investigation of the effects of automated vehicle technology. For instance, *(5)* and *(4)* used web-based survey tools to determine the willingness-to-pay of population in the U.S. for automated vehicles. An important dimension of disruptive technology like automated vehicles is that it doesn't have a strong reference in our daily lives for respondents to associate it with. Moreover, the respondents have not experienced such technology in any form yet. A web-based survey does not have the required features for a SP researcher to design choice scenario that best represent the reality and also help the respondents to completely understand them. It is further evident from the high variance in the willingness-to-pay reported in two studies that the web-based survey may not have provided the level of information necessary to respond objectively. McFadden in his keynote speech at the 2015 Conference of International Association of Travel Behaviour Research (IATBR) alluded on this shortcoming in the context of disruptive technologies and suggested the use of immersive simulators that can give the respondents near-real experience.

In recent years, the technological improvements in vehicle automation has been experiencing rapid advances, and several technology providers and auto manufacturers plan to introduce an automated self-driving vehicle by or before 2020 *(6)*. These developments will have significant impacts on the transportation system such as roadway safety, traffic congestion, environmental impacts, personal mobility, and travel behavior in general *(7)*. Connected and autonomous vehicles can offer greater safety and efficiency within the transportation system and potentially reduce a high proportion of the 90% of crashes that result from driver error *(8)*.

With the increasing advancement in the vehicle technology over the past decades, the future of car travel has become harder to predict. That said, although there are many researchers studying hybrid-electric vehicles, plug-in electric vehicles, car-sharing services, and on-demand services (Uber, Lyft, etc.), the real demand for the emerging CAVs are still under investigation. Recent studies have initiated the evaluation of the real impact of CAVs on reduction of crash rates, safety and security, its economic effects, etc. *(9, 10)*. However, understanding the future of this emerging mode of transportation is still a real challenge *(5, 10)*, and one of the reasons is precisely the lack of real setting and real experience that allow individuals to form preferences for CAV and researchers to measure these preferences.

Over the years, CAV demand forecasting has been studied by several researches, private enterprises, and industry affiliations, and have estimated AV adoption rates' predictions *(11, 12, 13, 14, 15)*. However, the predictions are based on extrapolating trends from older methods (e.g.

conventional vehicle technologies, expert opinions, or forecasts of supply-side variables) which cast an assumption of consistency between the two different vehicle types, and therefore is not a reliable method given the different nature of CAVs. More reliable demand forecasting requires AV specific factors such as willingness to pay (WTP), government's regulation technology adoption, effects of different levels of AV (e.g. level 1 and 2 automation technologies) such as lane centering assistance and adaptive cruise control. An AV forecast by Litman *(13)* made use of deployment and adoption of previous smart vehicle technologies (e.g. automatic transmission and hybrid-electric drive), and predicted that by 2040, 50% of vehicle sales, and 40% of all vehicle will be CAVs. Another forecasting study by *(12)* predicted that by 2030, the market size for level 2 and 3 automation technologies will account for up to $87 billion. However, they argue that level 4 technology is likely to be emerging by that time and level 3 automation will still be a premium option, which is expected account for only 8% of new car sales. Some predictions have gone as far as to forecast level 4 AV usage to be 90% of all trips. However, the experience with a seemingly new technology (all predictions of the electric vehicles markets proved to be wrong) should warn about the risks associated to the predictions of the market penetration of technologies that are still in evolution and that people have not experience with. To understand individuals' opinions about CAVs, *(16)* conducted a survey of 467 respondents, 30% of which were willing to spend more than $5000 to adopt full automation in their next vehicle purchase and around the same proportion of respondents showed interest in adopting AV technology, four years after its introduction on the market. 82% of respondents ranked safety as the most important factor affecting their adoption of CAVs, 12% and 6% ranked legislation and cost respectively as their main reason for choosing CAVs. Another survey by Underwood considered the opinions of 217 experts. His findings reported legal liability as the most difficult barrier to the wide implementation of level 5 CAVs, and consumer acceptance is the least. Additionally, approximately 72% of the experts suggested that CAVs should be at least twice as safe as the conventional vehicles before they are authorized for public use, while 55% of the experts indicated that level 3 CAVs are not practical because drivers may not take required actions in time *(17)*. In contrast to Schoettle and Sivak's findings, females are more interested in AV technologies compared to males who were more concerned about being forced to follow speed limits *(10)*. Focusing on different age groups, older participants (60 years and older) and younger participants (21-34 year olds) expressed the highest WTP in order to obtain self-driving technologies. As an important step in planning, a demand estimation and analysis need to be developed to evaluate the impact of the large-scale arrival of CAVs in the car market. The first step to any demand modeling is data collection which in this case faces several problems due to the very low number of CAV users. To address this issue, mainly web-based questionnaire about opinions and individuals' preferences for the adoption of these emerging vehicles *(5)*.

Daziano et al. used stated preferences (SP) surveys where the level of automation was one of the attributes that described the hypothetical choice situations *(4)*. As discussed before, while powerful and necessary in case of new contexts, SP surveys have been questioned for the lack of realism. The use of visual attributes (images) has been tested in the literature as a tool to improve realism (see a recent work by Patterson et al. *(2)*). Visualizations also play an important role in communicating transportation plans and policies. However, the use of visual images in a standard SP setting does not solve another major problem associated with using SP to study preferences for CAVs i.e. the lack of experience and hence of knowledge of the product, which has a major impact in the estimation of individuals' preferences *(18, 19)*. Recent advances in computer graphics and technology have provided new opportunities for generating more realistic virtual scenarios that are suitable for behavioural studies *(2)*. An emerging interactive technology in the field of transportation studies are Virtual Reality (VR) tools which have opened a new window for practical applications and scientific investigation of human perception and behaviour. VR allows the user to immerse in a set environment for in-depth evaluation of user perception and behaviour. Its ability to overlay the physical environment with virtual elements such as information or images, which can interact with the physical environment in real time, provides new possibilities for content delivery. VR allows the sensation of immersion in the activities on the screen and the virtual elements *(20, 21, 22, 23)*. The approach proposed in this paper consists in conducting a SP experiment within a VR environment. This approach is better than a typical SP experiment in terms of realism, because it puts respondents in a realistic environment and allows respondents to interact with and move within the environment as they would do in real life. Given the role played by realism in SP experiment, the level of realism achieved with VR environment represents a significant improvement. Other than that, our SP experiment implemented within a VR environment allows respondents to get direct experience with the product, which is the biggest single drawback of using standard SP to measure preferences for autonomous vehicles. VR environment experiments have successfully been conducted in various fields of cognitive studies. For example, studies have shown that people can develop realistic spatial knowledge in the VR environment that is similar to actual physical environments *(24, 25, 26)*. The main advantage of adapting VR in research studies is the freedom and versatility in setting up experiments which enables scientists to measure physical reactions of participants by adopting electrocardiography, skin conductance, electroencephalogy, and eye-tracking enables researchers to understand visual perception better by monitoring gaze behaviour *(2, 27)*.

Major benefits of using virtual environment is the ability to examine various scenarios for incidents, employ appropriate treatments/policies and analyse their performances before making irreversible decisions that are either difficult or expensive in the real word. An Immersive Head Mounted Virtual Reality environment (IHMVR) is a VR display device that uses an optical system to directly present virtual scenes received by the display and works with the human brain to produce a strong sense of immersion *(28)*. Immersive virtual environments which allow a locomotive interface may preserve the perception-action coupling that is critical in examining many visual timing skills *(29)*. Researchers also applied virtual environments to study pedestrian route choice and reaction to information in evacuation scenarios. Sobhani et al. studied the effects of multi-tasking (walking and using smartphone) on adults' street crossing performance by using a simulated intersection crossing task programmed in an IHMVR with an integrated audio and sensors which allowed close emulation of real word behaviour of pedestrians *(30)*. Using a high-fidelity street crossing simulator, Neider et al. showed that naturalistic cell phone conversations impair crossing performance and increase crash rates *(31)*. Similarly, Stavrinos et al. demonstrated significant costs to simulated crossing performance while conversing on a cell phone *(32, 33)*.

**VIRTUAL IMMERSIVE REALITY ENVIRONMENT (VIRE)**

Leveraging upon the technological developments in Head Mounted Display (HMD) and virtual reality, VIRE aim at bringing realism, interactivity, immersiveness into the choice scenario. Furthermore, VIRE enables us to design complex experiments to directly study phenomena like mental processes, without putting extra burden on the respondents. In the standard setting, the respondents can experience forthcoming transportation technologies and services very realistically, by projection of the traffic simulation directly to their eyes and real-time interactions between them and simulation objects (e.g. CAV, cyclists, etc.). This also improves the understanding of the scenario by respondents in comparison to conventional techniques e.g. text or visuals based stated preference experiments. As pointed out in Figure 1, VIRE has four main systems that interacts with each other and the respondents. These systems are a mix of software and hardware components working together.

**Scenario Development (SD) System**

The theoretical experiment design is translated into a 3D scenario using Unity, an open-source gaming engine *(34)*. The scenario is created using 3D models of the buildings, vehicles, persons, cyclists, trees, roads, and other components necessary. This system is capable of importing already available 3D resources and incorporating them into the scenario. The road network internally is coded as a collection of directed links and nodes. VIRE is capable of importing any road network from the freely available OpenStreetMaps to VIRE road network.

**Multi-Modal Traffic Micro-Simulation (MMTM) System**

VIRE currently supports the large-scale simulation of vehicles, person, and cyclists. Vehicular movement is simulated using the Intelligent Driver Model (IDM), which is a variant of widely used car following model *(35)*. IDM allows us to incorporate the behaviour of autonomous vehicles–especially, avoiding collisions with pedestrians and maintaining a safe stopping distance from them. In the VIRE implementation, we used the IDM parameters calibrated by Kenting and Treiber *(35)*. For simulating the pedestrian movement, we adapted the social force model *(36)* to incorporate the multi-modal nature of the simulation. Pedestrians not only calculates the forces with respect to other pedestrians and obstacles, but also the approaching vehicles and cyclists. In the VIRE implementation, we used the social force parameters calibrated by Sahaleh et al. *(36)*. For cyclist's movement, to best of our knowledge there is no dedicated flow model been developed. Thus, we adapted the social force model to mimic the cyclist's movement. Please also note that the currently implemented models and their parameters within VIRE can be very easily changed to specific requirements of the experiments. The simulator is implemented in $C^{\#}$ language in order to seamlessly interact with Unity, which is also implemented in the same development environment. This system uses the network, resources, and environment developed by the SD system to simulate the multi-modal traffic flow.

**Virtual Environment Projection (VEP) System**

Using the 3D visualization and animation support available in Unity gaming engine, this system projects the positions and interactions of various dynamic objects (i.e. vehicles, pedestrians, and cyclists) onto a commercially available Head Mounted Display. Currently, we are only using Oculus Rift for projection, but our projection system can support other HMDs including HTC Vive, Microsoft HoloLens, and others. The projection system takes the environment with static 3D objects that is created by SD system and overlay the dynamic moving objects, whose positions are provided by the MMTM system. VIRE is also capable of generating and associating audios to 3D objects, for example a car passing by will make the appropriate noise that the respondent can hear. High importance is given to the scene refresh rate in order to create a realistic environment. Current HMDs are capable of refreshing at 90Hz. Thus, we designed our traffic simulation and scene rendering in such a way that more than 10,000 agents can be simulated while maintaining at least 90Hz. This is achieved by employing parallel processing and localized rendering. We are currently using the Core i7 6700K (4.00GHz, 8M Cache, Quad-Core) processor and Nvidia GeForce GTX 1080 graphics card to ensure maximum processing capabilities.

**Response Tracking (RT) System**

The distinctive feature of VIRE is that it cannot only capture various types of responses from the respondent but can also let them interact with other virtual objects (Figure 2). Currently, we can record the exact trajectory of the respondent moving within an area of 4.5m x 2m using the Oculus Motion Sensors. This area can be extended if required, by using appropriate hardware with minimal changes in the core software. VIRE can also record any vocal responses (i.e. the Ohs and Ahs). The video of actual movement of the respondent can be recorded. Moreover, VIRE also records which direction the respondents are looking at by recording the orientation of their head. All these information about the respondent are transmitted to the MMTM system, which updates the simulation according the responses and interactions with virtual objects. Additionally, we have implemented a wrapper around Oculus Touch hardware, so that it can be converted into a hand-held device. This can be useful to examine the respondent's behaviour in terms of smartphone use, driving, etc. This functionality can be extended to any other hand-held device. Sobhani et al. *(30)* is an example of the use of this feature where we studied the behaviour of smartphone distracted pedestrian crossing the road.

**EXPERIMENT DESIGN**

To demonstrate the distinctive functionalities of VIRE system, a controlled SP like experiment related to the preference of autonomous vehicles in dense urban environment was designed. A pedestrian dominant area outside Laurier metro station in Montréal was selected (Figure 3). The intersection of Laurier and Rivard streets was created in VIRE. The controlled experiment involved two scenarios:

a) Current state: signalized intersection with vehicles driven by people (Figure 3(a))
b) Hypothetical scenario: unsignalized intersection with autonomous vehicles and pedestrian priority.

Both scenarios were developed in VIRE. In case of scenario a, pedestrians would wait for the walk signal to turn green. They could jay-walk but were not allowed to run. Running was prohibited because the respondent may hit some physical obstacle in the laboratory or trip themselves with the wires. Furthermore, if they jay-walk and a vehicle was coming, the safe distance may not have been maintained. In that case, the respondents were notified that an accident has happened. In case of scenario b, autonomous vehicles would stop at a safe distance if the respondents try to enter the crossing. Here too, the respondents were not allowed to run. Based on these two scenarios, we investigated the following questions:

a) Analysis of the pedestrian acceptance of autonomous vehicles and associated infrastructure changes in urban settings using VIRE
b) Comparison of VIRE with existing experiment tools
    - Text only
    - Visuals

Respondents were first given a piece of paper that described the two scenarios using text only. They were then asked to choose their preference among the two. In the second part of the experiment, they were shown an animation of the two scenarios. The animation was repeated several times for a respondent to understand what is going on and familiarize with the scenario. Their preference was asked again. In the third part of the experiment, they were made to wear the HMD and physically walk through the two scenarios developed in VIRE. This part of the experiment was repeated 10 times with different arrival times and gaps between the vehicles that were precomputed. Every respondent would go through the 10 trials in the same order. The pattern was controlled by a Poisson arrival distribution (μ = 4sec). The horizon of each trial was 2 mins. If the respondent was not able to cross during this time, the trial would stop. The speed limit was set of 50 km/hr. In each trial, based on safety literature, there were two "safe gaps" randomly inserted between the cars–one was 5 sec and other was 7 sec. This gap was such that if the pedestrian crosses, the incoming car doesn't need to stop. Respondent's trajectory in the virtual environment, head movement, vocal and physical expressions were also recorded. During the experiments, respondents were constantly inquired if they felt uncomfortable or symptoms of motion sickness. We were to stop the experiment immediately if the respondent replied positively. However, none of the participant experienced the motion sickness.

In total 42 respondents in Montréal and Toronto participated in these experiments. Our sample included university students older than 18 years, both male and female. Basic socio-demographics and information on their dominant mode and activity patterns were collected. Their level of familiarity with the virtual reality was also recorded. In terms of mode, 12% reported that their primary mode was biking, 45% public transit, 19% walk, and 24% car.

**RESULTS AND ANALYSIS**

Among the respondent, 43% were female and 57% male. Average age of the respondents was 26 years, with minimum of 19 years, maximum 40 years, and a standard deviation of 5 years. 41.5% respondents were residents of Toronto and 58.5% were from Montréal. 80% of the respondents had driving license. 36% of the respondents reported that they have experienced the HMD based virtual reality before.

Figure 4 shows the comparative analysis of the choice preferences using the three experiment tools i.e. text-only, visual, and virtual immersive reality environment. In case of text-only (Figure 4(a)), 50% of the respondents reported that they would prefer the scenario where autonomous vehicles are operating on urban streets without signalization and with pedestrian priority. In case of simple visualization this percentage decreased below 40%. We hypothesize that the reason behind this is that the information provided by second tool (visual animation) doesn't add any value in terms of scenario realism. Instead, it affects the respondent's perception of autonomous vehicles in a negative way. Using VIRE, the realism of scenario is improved resulting in a near-realistic, immersive, and interactive experience for respondents. Thus, the preference for autonomous vehicle scenario increases to 70%.

Clear differences among the genders are noticeable (Figure 4(b)), with female respondents being more conservative about the autonomous vehicles than male in all three experimental tools. In the case of all three tools, males are more likely to prefer the scenario with autonomous vehicles, while this preference increases when using the VIRE. Female respondents prefer (about 70%) the current state (i.e. normal vehicles and signalized intersection) when the experiment is run with text-only or simple visualization. However, this percentage flips when the female respondents experience the scenario in VIRE. There are some notable differences between respondents from Toronto and Montréal (Figure 4(c)) as well. In general, respondents in Toronto were more conservative about autonomous vehicles while using VIRE as the experiment tool. This can be attributed to the general differences between the urban streets of Toronto and Montréal. Toronto is a typical North-American city with wider roads, higher volume of traffic and speed limits, and longer block-sizes. Central Montréal has a more European feel, with narrower roads, lower speed limits, smaller blocks, and more mixed-use areas. However, it should be noted that due to the small sample size, we cannot generalize this observation.

To quantitatively assess the factors affecting the respondent's preferences and differences between the tool used, we developed three multinomial Logit models. Table 1 reports the three models with parameter values that were found to be statistically significant. We also jointly estimated the three models by fixing the scale parameter ($\mu = 1$) for VIRE model and estimating the scale for the other two. The estimated scale parameter for MNL based on text-only and basic visualization was found to be 1.16, which indicates differences in the samples. The model developed from VIRE has better overall statistics with lowest log-likelihood. All the individual parameters in VIRE based MNL had right expected signs. In terms of parameter values, the respondents with higher age (millennials) preferred more the autonomous vehicle. Compared to male, female had a negative preference towards autonomous vehicles. Male whose primary mode of travel was biking preferred more the autonomous vehicle. This may be due to the fact that they expect the AVs to always follow the rules and thus they feel safer. Torontonians compared to Montréalers had a lesser preference. Furthermore, we found that the prior experience with HMD has a positive and significant effect towards the choice of AVs with unsignalized intersection. This may be due to the fact that in general they felt more comfortable using the HMD and were able to absorb more information in the scene.

**CONCLUDING REMARKS**

We present Virtual Immersive Reality Environment (VIRE), a generic platform for developing travel related stated preference experiments that are highly realistic, interactive, immersive, and can efficiently collect information beyond just preferences. We demonstrate the use of VIRE in the pedestrian preferences related to autonomous vehicles and the related infrastructure changes. Pedestrian can physically walk in highly realistic virtual 3D model of a real intersection in Montréal. The data collected using VIRE is compared with the predominantly used tools i.e. text based and visual animation. We found that the understanding of respondents increased, and their preferences became more consistent. The joint multinomial Logit model estimated also showed the difference in scale parameter for VIRE and other tools. In the VIRE experiment we controlled for the arrival and gap of the vehicles. We plan to include that in the model to understand the effects of level of autonomous vehicle traffic and gaps on the pedestrian preferences.

SP experiments have been known to suffer from the lack of realism. This issue becomes even more relevant when the transportation technology or service is highly innovative and without a well understood reference e.g. autonomous vehicles and their operations on the urban roads. VIRE not only addresses the realism issues but also go beyond by collecting additional information on the movement, vocal responses (i.e. the Ohs and Ahs), direction in which respondent is looking, and interactions with other objects, without any additional burden on the respondent. This information can help us develop more complex and realistic models of behaviour. One such application could be the explicit modelling of the mental processes while deciding on choice preference. Our next step is to design and develop SP experiments in VIRE that can let us investigate this research.

In terms of functionality, we are currently working on incorporating a manual driving hardware (steering wheel, break, and acceleration) in VIRE, in order to develop experiments that compare the experience of driving manually versus being driven in autonomous vehicles. In near future, we plan to enhance VIRE so that it can support multiple respondents in one experiments. This functionality can be very useful in studying group behaviour e.g. two friends walking together or a family driving in an autonomous vehicle.

One of the biggest hurdle in the mainstream use of virtual and augmented reality (VR/AR) in travel related SP experiments have been the amount of software engineering, traffic flow theory, and simulation related knowledge and effort required to develop the data collection tool. In near future, we plan on bringing down this entrance barrier by sharing the VIRE source code with the transportation research community. We envision that this will extensively enhance our collective ability to use VR/AR for understanding travel behaviour, its drivers, and the influence of new technology and services. We plan to conduct this experiment on a larger sample involving 300-600 participants. A larger dataset will allow us to develop richer models and test our hypotheses more extensively.

**DECLERATION**

The authors confirm contribution to the paper as follows: study conception and design: Farooq, Cherchi, Sobhani; data collection: Farooq, Sobhani; analysis and interpretation of results: Farooq, Cherchi; draft manuscript preparation: Farooq. All authors reviewed the results and approved the final version of the manuscript.

**REFERENCES**

1. Cherchi, E. and D. A. Hensher (2015) Workshop synthesis: Stated preference surveys and experimental design, an audit of the journey so far and future research perspectives, Transportation Research Procedia, 11, 154–164.
2. Patterson, Z., J. M. Darbani, A. Rezaei, J. Zacharias and A. Yazdizadeh (2017) Comparing text-only and virtual reality discrete choice experiments of neighbourhood choice, Landscape and Urban Planning, 157, 63–74.
3. Fagnant, D. J. and K. M. Kockelman (2014) The travel and environmental implications of shared autonomous vehicles, using agent-based model scenarios, Transportation Research Part C: Emerging Technologies, 40, 1–13.
4. Daziano, R. A., M. Sarrias and B. Leard (2017) Are consumers willing to pay to let cars drive for them? analyzing response to autonomous vehicles, Transportation Research Part C: Emerging Technologies, 78, 150–164.
5. Bansal, P., K. M. Kockelman and A. Singh (2016) Assessing public opinions of and interest in new vehicle technologies: an Austin perspective, Transportation Research Part C: Emerging Technologies, 67, 1–14.
6. Bierstedt, J., A. Gooze, C. Gray, J. Peterman, L. Raykin and J. Walters (2014) Effects of next-generation vehicles on travel demand and highway capacity, FP Think Working Group, 10–11.
7. LaMondia, J. J., D. J. Fagnant, H. Qu, J. Barrett and K. Kockelman (2016) Long-distance travel mode shifts due to automated vehicles: A statewide mode-shift simulation experiment and travel survey analysis, paper presented at the Transportation Research Board 95th Annual Meeting, no. 16-3905.
8. Bellis, E. and J. Page (2008) National motor vehicle crash causation survey (nmvccs) sas analytical users manual, Technical Report.
9. Fagnant, D. J. and K. Kockelman (2015) Preparing a nation for autonomous vehicles: opportunities, barriers and policy recommendations, Transportation Research Part A: Policy and Practice, 77, 167–181.
10. Schoettle, B. and M. Sivak (2014) A survey of public opinion about autonomous and self-driving vehicles in the USA, the UK, and Australia.
11. Hars, A. (2014) Autonomous vehicle roadmap: 2015-2030, http://www.driverless-future.com/?p=678.
12. Laslau, C., M. Holman, M. Saenko, K. See and Z. Zhang (2014) Set autopilot for profits: Capitalizing on the $87 billion self-driving car opportunity, Lux Research.
13. Litman, T. (2014) Autonomous vehicle implementation predictions, Victoria Transport Policy Institute, 28.
14. Mosquet, X., T. Dauner, N. Lang, N. Rubmann, A. Mei-Pochtler, R. Agrawal and F. Schmieg (2015) Revolution in the driver's seat: The road to autonomous vehicles, Boston Consulting Group, 11.
15. Rowe, R. (2015) Self-Driving Cars, Timeline, http://www.topspeed.com/cars/car-news/self-driving-cars-timeline-ar169802.html.
16. Casley, S., A. Jardim and A. Quartulli (2013) A study of public acceptance of autonomous cars, Worcester Polytechnic Institute, Bachelor Thesis.
17. Underwood, S. (2014) Automated vehicles forecast vehicle symposium opinion survey, paper presented at the automated vehicles symposium.

18. Jensen, A. F., E. Cherchi and S. L. Mabit (2013) On the stability of preferences and attitudes before and after experiencing an electric vehicle, Transportation Research Part D: Transport and Environment, 25, 24–32.
19. Cherchi, E. (2016) A stated choice experiment to study the effect of social conformity in the preference for electric vehicles, Technical Report, Newcastle University.
20. Animesh, A., A. Pinsonneault, S.-B. Yang and W. Oh (2011) An odyssey into virtual worlds: Exploring the impacts of technological and spatial environments on intention to purchase virtual products, MIS Q., 35 (3) 789–810, ISSN 0276-7783.
21. Faiola, A., C. Newlon, M. Pfaff and O. Smyslova (2013) Correlating the effects of flow and telepresence in virtual worlds: Enhancing our understanding of user behavior in game-based learning, Computers in Human Behavior, 29 (3) 1113–1121.
22. Jennett, C., A. L. Cox, P. Cairns, S. Dhoparee, A. Epps, T. Tijs and A. Walton (2008) Measuring and defining the experience of immersion in games, International journal of human-computer studies, 66 (9) 641–661.
23. Nah, F. F.-H., B. Eschenbrenner and D. DeWester (2011) Enhancing brand equity through flow and telepresence: A comparison of 2d and 3d virtual worlds, MIs Quarterly, 731–747.
24. O'Neill, M. J. (1992) Effects of familiarity and plan complexity on wayfinding in simulated buildings, Journal of Environmental Psychology, 12 (4) 319–327.
25. Ruddle, R., S. Payne and D. Jones (1997) Navigating buildings in virtual environments Experimental investigations using extended navigational experience, Journal of Experimental Psychology: Applied, 3 (2) 143–159.
26. Tlauka, M. and P. N. Wilson (1996) Orientation-free representations from navigation through a computer-simulated environment, Environment and Behavior, 28 (5) 647–664.
27. Wiener, J. M., C. Hölscher, S. Büchner and L. Konieczny (2012) Gaze behaviour during space perception and spatial decision making, Psychological research, 76 (6) 713–729.
28. Wang, Y., W. Liu, X. Meng, H. Fu, D. Zhang, Y. Kang, R. Feng, Z. Wei, X. Zhu and G. Jiang (2016) Development of an immersive virtual reality head-mounted display with high performance, Applied Optics, 55 (25) 6969–6977.
29. Wu, H., D. H. Ashmead and B. Bodenheimer (2009) Using immersive virtual reality to evaluate pedestrian street crossing decisions at a roundabout, paper presented at the Proceedings of the 6th Symposium on Applied Perception in Graphics and Visualization, 35–40.
30. Sobhani, A., B. Farooq and Z. Zhong (2017) Immersive head mounted virtual reality based safety analysis of smartphone distracted pedestrians at street crossing, paper presented at the Road Safety and Simulation conference, no. 1-12.
31. Neider, M. B., J. S. McCarley, J. A. Crowell, H. Kaczmarski and A. F. Kramer (2010) Pedestrians, vehicles, and cell phones, Accident Analysis & Prevention, 42 (2) 589–594.
32. Stavrinos, D., K. W. Byington and D. C. Schwebel (2009) Effect of cell phone distraction on pediatric pedestrian injury risk, Pediatrics, 123 (2) 179–185.
33. Stavrinos, D., K. W. Byington and D. C. Schwebel (2011) Distracted walking: cell phones increase injury risk for college pedestrians, Journal of safety research, 42 (2) 101–107.
34. Unity developers (2017) Unity user manual (2017.1). https://docs.unity3d.com/Manual/index.html.
35. Kesting, A. and M. Treiber (2008) Calibrating car-following models by using trajectory data: Methodological study, Transportation Research Record: Journal of the Transportation Research Board, (2088) 148–156.

36. Sahaleh, S., M. Bierlaire, B. Farooq, A. Danalet and F. Hänseler (2012) Scenario analysis of pedestrian flow in public spaces, paper presented at the 12th Swiss Transport Research Conference, no. EPFL-CONF-181239.

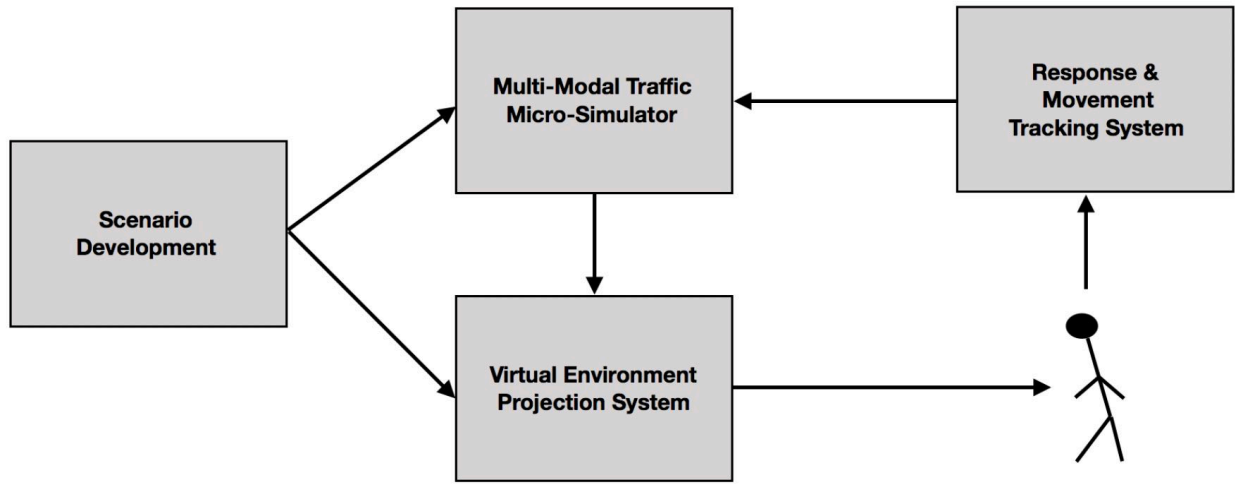

FIGURE 1 Components and interactions in VIRE

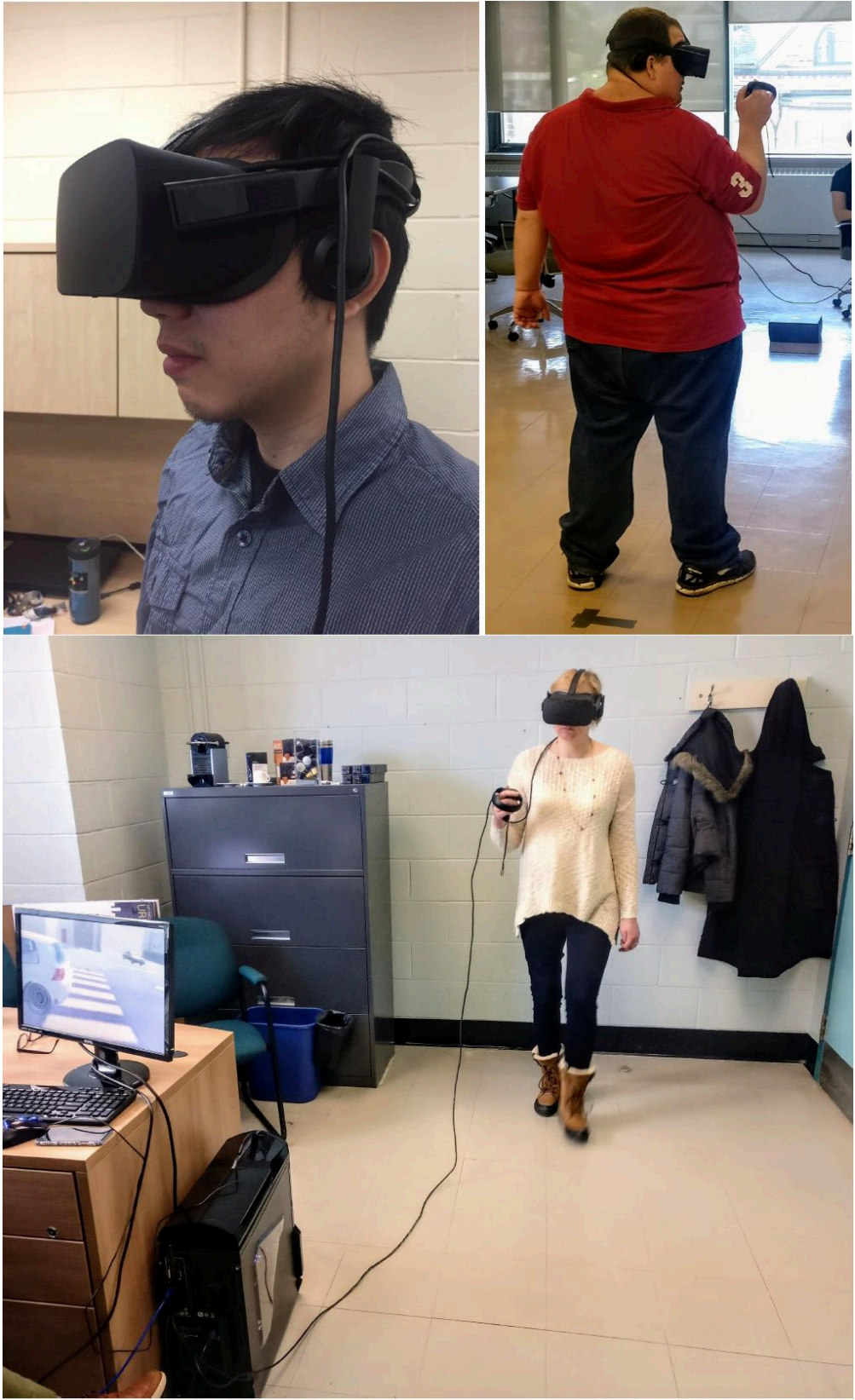

FIGURE 2 Respondents experiencing VIRE system

Farooq, Cherchi, and Sobhani 17

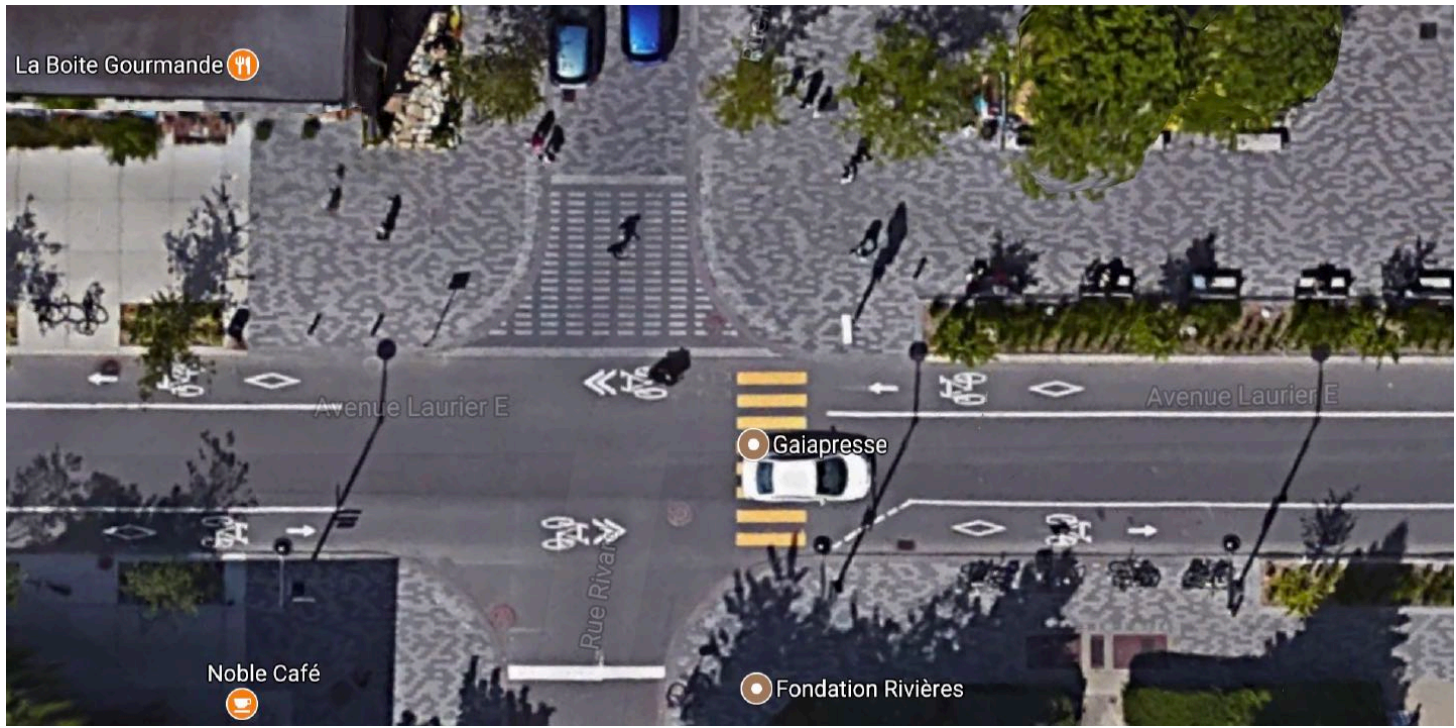

(a) Google satellite view

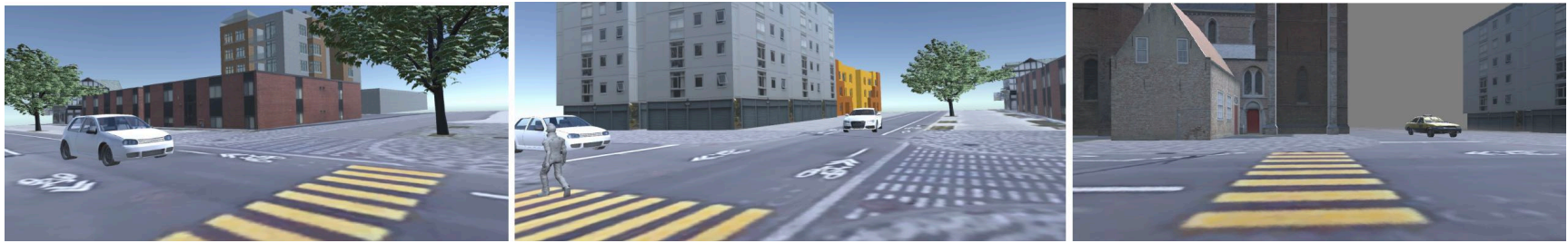

(b) Virtual representation of the intersection

FIGURE 3 Intersection setup for the experiment

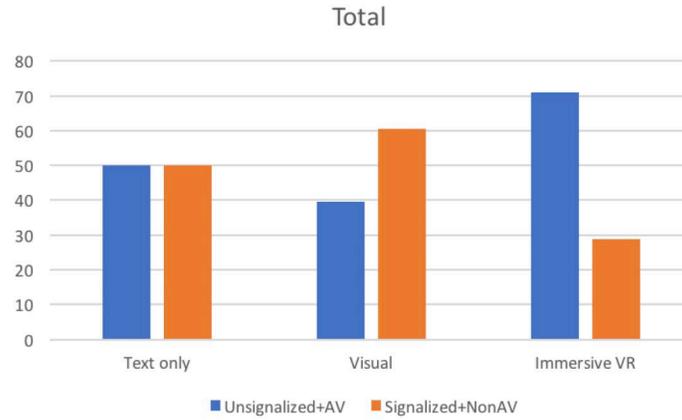
(a) Overall choice behaviour

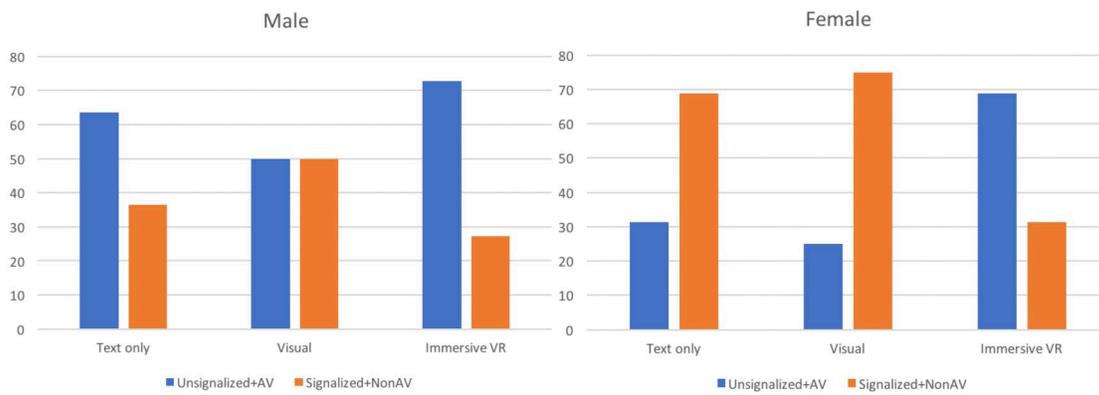
(b) Choice behaviour by gender

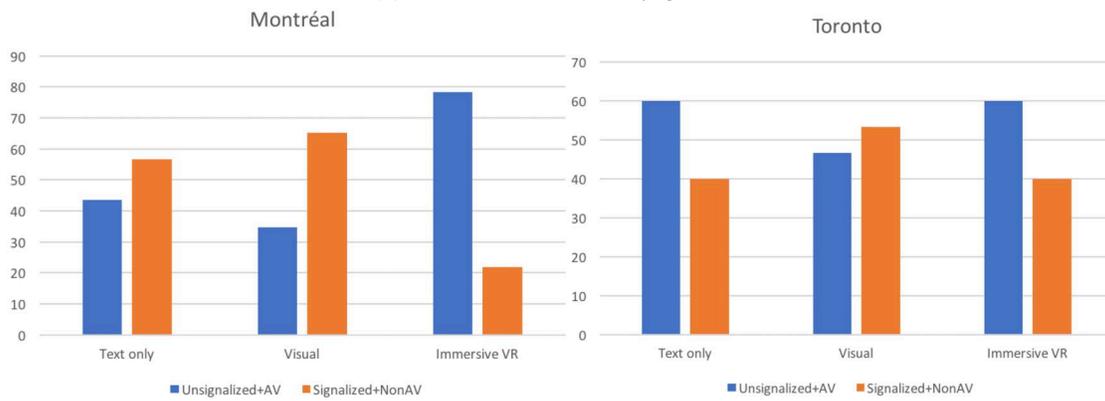
(c) Choice comparison by city

FIGURE 4 Choice comparison between text-Only, visual, and VIRE

Farooq, Cherchi, and Sobhani 19TABLE 1 Multinomial Logit model estimates

| Parameter | Text | | Visualization | | VIRE | |
|---|---|---|---|---|---|---|
| | value | st-err | value | st-err | value | st-err |
| $ASC_{AV}$ | 1.90 | 1.87 | -2.12 | -1.88 | -3.49 | -1.64 |
| $\beta_{Age}$ | 0.69 | 0.69 | 0.65 | 0.63 | 1.76 | 0.83 |
| $\beta_{Female}$ | -1.61 | 0.80 | -1.42 | 0.82 | -1.14 | -1.32 |
| $\beta_{Bike,male}$ | 0.972 | 1.72 | 0.212 | 1.12 | 12.70 | 0.922 |
| $\beta_{Toronto}$ | -1.09 | 0.75 | -0.82 | 0.74 | -1.18 | -0.82 |
| $\beta_{HMD}$ | - | - | - | - | 2.03 | 0.84 |
| $\mathcal{L}$ | -25.28 | | -25.18 | | -18.98 | |
| $\rho^2$ | 0.132 | | 0.135 | | 0.348 | |
| Obs. | 42 | | 42 | | 42 | |